\documentclass[aps,prd,groupedaddress,showpacs,showkeys]{revtex4}
\usepackage{latexsym,epsfig,amsmath,amssymb}
\usepackage{amssymb}
\usepackage{amsmath}
\usepackage{amsfonts}
\usepackage{epsfig} 
\usepackage{verbatim} 

\newcommand{\seq}{\begin{subequations}}
\newcommand{\sen}{\end{subequations}}
\newcommand{\eq}{\begin{eqnarray}}
\newcommand{\en}{\end{eqnarray}}
\newcommand{\ra}{\rangle}
\newcommand{\la}{\langle}

\def\dppm{D^\pm}
\def\dpp{D^+}
\def\dpm{D^-}
\def\dpmps{D^{\ast \mp}}
\def\dpps{D^{\ast +}}
\def\dpms{D^{\ast -}}

\def\d{D^0}
\def\db{\bar{D}^{0}}
\def\ds{D^{\ast  0}}

\def\dbs{\bar{D}^{\ast 0}}

\def\L2{\Lambda^2}

\def\jp{J/\psi} 
\def\jpsi{J_\psi}

\begin{document}

\title{$X(3872)$ as a hadronic molecule and its
decays to charmonium states and pions} 

\noindent
\author{
        Yubing Dong$^{1,2,3}$, 
        Amand  Faessler$^1$,   
        Thomas Gutsche$^1$, 
        Sergey Kovalenko$^4$,        
        Valery E. Lyubovitskij$^1$\footnote{On leave of absence
        from Department of Physics, Tomsk State University,
        634050 Tomsk, Russia}
\vspace*{1.2\baselineskip}}

\affiliation{$^1$ Institut f\"ur Theoretische Physik,  
Universit\"at T\"ubingen,\\
Kepler Center for Astro and Particle Physics, \\ 
Auf der Morgenstelle 14, D--72076 T\"ubingen, Germany
\vspace*{1.2\baselineskip} \\
$^2$ Institute of High Energy Physics, Beijing 100049, P. R. China 
\vspace*{1.2\baselineskip} \\ 
$^3$ Theoretical Physics Center for Science Facilities (TPCSF), CAS, 
Beijing 100049, P. R. China 
\vspace*{1.2\baselineskip} \\ 
$^4$ Centro de Estudios Subat\'omicos(CES),
Universidad T\'ecnica Federico Santa Mar\'\i a, \\
Casilla 110-V, Valpara\'\i so, Chile\\} 

\date{\today}

\begin{abstract} 

The $X(3872)$ with quantum numbers $J^{PC} = 1^{++}$ 
is considered as a composite hadronic state 
comprised of the dominant molecular $\d\ds$ component and other
hadronic pairs -- $D^{\pm} D^{\ast\,\mp}$, 
$\jp\omega$ and $\jp\rho$. Applying the compositeness condition 
we constrain the couplings of the $X(3872)$ to its constituents. 
We calculate two- and three-body hadronic decays of the $X(3872)$ 
to charmonium states $\chi_{cJ}$ and pions using a phenomenological 
Lagrangian approach. Next using the estimated $X\jp\omega$ and $X\jp\rho$ 
couplings we calculate the widths of $X(3872) \to \jp + h$ transitions, 
where $h$ = $\pi^+\pi^-$, $\pi^+\pi^-\pi^0$, $\pi^0\gamma$ and $\gamma$.  
The obtained results for the decay pattern of the $X(3872)$ in
a molecular interpretation
could be useful for running and planned experiments. 

\end{abstract}



\maketitle

\newpage

\section{Introduction}

The $X(3872)$ is one of the new meson resonances discovered during the last 
years~\cite{Amsler:2008zz}, whose properties cannot be simply explained 
and understood in conventional quark models. Several structure 
interpretations for the $X(3872)$ have been proposed in the literature 
(for a status report see e.g. 
Refs.~\cite{Swanson:2006st,Bauer:2005yu,Voloshin:2007dx}). 
In the context of molecular 
approaches~\cite{Voloshin:1976ap}-\cite{Fleming:2008yn} the $X(3872)$ 
can be identified with a weakly bound hadronic molecule whose constituents 
are $D$ and $D^\ast$ mesons. The reason for this natural interpretation is 
that its mass $m_X$ is very close to the $\d \dbs$ threshold and hence 
is in analogy to the deuteron --- a weakly bound state of proton and neutron. 
Note, that the idea to treat the hidden charm states as hadronic molecules 
traces back to Refs.~\cite{Voloshin:1976ap,DeRujula:1976qd}. 
Originally it was proposed that the state $X(3872)$ is a superposition 
of $\d \dbs$ and $\db \ds$ pairs. Later (see e.g. discussion 
in Refs.~\cite{Swanson:2003tb,Voloshin:2004mh,Braaten:2005ai}) also
other structures, such as a charmonium or even other
meson pair configurations, were discussed in addition to the 
$\d  \dbs +$ charge conjugate (c.c.) component 
(here and throughout the paper we use the convention that 
$\dbs$ does not change sign under charge 
conjugation. See detailed discussion in Ref.~\cite{Thomas:2008ja}).  
The possibility of two nearly degenerated $X(3872)$ states with positive 
and negative charge parity has been discussed in  
Refs.~\cite{Terasaki:2007uv,Gamermann:2007fi}. 

This paper focuses on the hadronic 
$X \to \chi_{cJ} + (\pi^0, 2\pi)$, 
$X \to \jp + (2\pi, 3\pi)$ and radiative 
$X \to \jp + (\pi^0\gamma, \gamma)$ decays.  
The $X(3872)$ with quantum numbers $J^{PC} = 1^{++}$ 
is considered as a composite hadronic state including a 
dominant molecular $\d\ds$ component and other
hadronic pairs -- $D^{\pm} D^{\ast\,\mp}$, 
$\jp\omega$ and $\jp\rho$. 
This idea was originally proposed 
in~\cite{Swanson:2003tb}. Applying the compositeness condition 
we constrain the couplings of $X(3872)$ to its constituents. 
We calculate two- and three-body hadronic decays of the $X(3872)$ 
to charmonium states $\chi_{cJ}$ and pions using a phenomenological 
Lagrangian approach. Next, using the estimated $X\jp\omega$ and $X\jp\rho$ 
couplings we calculate the widths of $X \to \jp + h$ transitions, 
where $h$ = $\pi^+\pi^-$, $\pi^+\pi^-\pi^0$, $\pi^0\gamma$ and $\gamma$.  
Present experimental numbers for the ratios of observed decay modes of 
$X(3872)$ by the Belle~\cite{Abe:2005ix} and 
BABAR~\cite{Aubert:2008rn} Collaborations are  
\eq\label{data_strong}
\displaystyle\frac{\Gamma(X \to \jp \pi^+ \pi^- \pi^0)}
{\Gamma(X \to \jp \pi^+ \pi^-)}= 1.0 \pm 0.4 (\text{stat})   
\pm 0.3 (\text{syst})   
\en 
and 
\eq\label{data_radiative}
\displaystyle\frac{\Gamma(X \to \jp\gamma)}
{\Gamma(X \to \jp\pi^+\pi^-)}= 0.14 \pm 0.05 \ \ (\text{Belle}); \ 
                             0.33 \pm 0.12 \ \ (\text{{\it BABAR}}) \,. 
\en 
The theoretical analysis of hadronic and radiative decays 
of $X(3872)$ has been carried out using a charmonium 
interpretation~\cite{Barnes:2003vb,Swanson:2004pp}, different molecular 
approaches~\cite{Braaten:2003he,Swanson:2004pp,%
Braaten:2005ai,Dong:2008gb,Fleming:2008yn} with possible inclusion 
of charmonium and other hadronic components in the $X$ wave 
function, QCD sum rules in~\cite{Navarra:2006nd}, multipole expansion 
in QCD and chiral properties of soft pions~\cite{Dubynskiy:2007tj}. 
In particular, pionic transitions from $X(3872)$ to the charmonium states 
$\chi_{cJ}$ have been considered using a pure charmonium and 
four-quark~\cite{Dubynskiy:2007tj} structure for the $X(3872)$ and later 
on in the molecular interpretation~\cite{Fleming:2008yn}. A conclusion was 
that the decay rates significantly depend on the structure interpretation 
of the $X(3872)$. It was also proposed that the $X(3872)$ 
to $\jp$ transitions are dominated by short--distance effects and 
in the mechanism of these transitions the $\jp\omega$ 
and $\jp\rho$ components of $X$ probably play the essential 
role~\cite{Braaten:2005jj,Braaten:2005ai}. 
 
In Refs.~\cite{Faessler:2007gv,Dong:2008gb,Branz:2008cb} 
we developed the formalism for the study of recently observed 
exotic meson states (like $D_{s0}^\ast(2317)$, $D_{s1}(2460)$, $X(3872)$, 
$\cdots$) as hadronic molecules. In Ref.~\cite{Dong:2008gb} 
we extended our formalism to the decay $X \to \jp \gamma$ assuming that the 
$X$ is the $S$--wave, positive charge parity $(\d  \dbs + \ds \db)/\sqrt{2}$ 
molecule. 
As for the case of the $D_{s0}^{\ast}$ and  $D_{s1}$ states, 
a composite (molecular) structure of the $X(3872)$ meson is defined 
by the compositeness condition $Z=0$~\cite{Weinberg:1962hj,%
Efimov:1993ei,Anikin:1995cf} 
(see also Refs.~\cite{Faessler:2007gv,Dong:2008gb,Branz:2008cb}).  
This condition 
implies that the renormalization constant of the hadron wave function 
is set equal to zero or that the hadron exists as a bound state of its 
constituents. The compositeness condition was originally 
applied to the study of the deuteron as a bound state of proton and
neutron~\cite{Weinberg:1962hj}. Then it was extensively used
in low--energy hadron phenomenology as the master equation for the
treatment of mesons and baryons as bound states of light and heavy
constituent quarks (see e.g. Refs.~\cite{Efimov:1993ei,Anikin:1995cf}). 
By constructing a phenomenological Lagrangian including the 
couplings of the bound state to its constituents and the constituents 
with other particles we calculated one--loop 
meson diagrams describing different decays of the molecular states 
(see details in~\cite{Faessler:2007gv,Dong:2008gb,Branz:2008cb}). 
In Ref.~\cite{Dong:2008gb} we estimated the role of a possible charmonium 
component in the $X(3872)$. We showed that the charmonium contribution to the 
$X \to \jp \gamma$ decay width is suppressed in comparison with  
the molecular $\d\ds$ component. As already stressed before, here we 
consider the $X(3872)$ as a superposition of the molecular $\d\ds$ component
and 
other hadronic pairs -- $D^{\pm} D^{\ast\,\mp}$, 
$\jp\omega$ and $\jp\rho$. 
Because of the dominance of the $\d\ds$ component 
in the transitions of $X$ into charmonium states $\chi_{cJ}$ and pions 
we estimate these decays using only that component. In the analysis 
of the decay widths with $\jp$ in the final state we will use the 
effective couplings $X\jp\omega$ and $X\jp\rho$ deduced from the 
compositeness condition. 

In the present paper we proceed as follows. In Sec.~II
we first discuss the basic notions of our approach. We discuss the effective
mesonic Lagrangian for the treatment of the $X(3872)$ meson 
as a superposition of the $\d \dbs + \ds \db$ molecular component with the
additional
$D^+ D^{\ast\, -} + D^- D^{\ast\, +}$ and $\jp\omega$ and 
$\jp\rho$ hadronic pairs. Second, we consider the two--body hadronic 
decays $X(3872) \to \chi_{cJ} + \pi^0 (2\pi^0)$.  
Third, we discuss decays with $\jp$ in the final state.   
In Sec.~III we present our numerical results and perform 
a comparison with other theoretical approaches. 
Finally, in Sec.~IV we present a short summary of our results.

\section{Approach} 

\subsection{Structure of the $X(3872)$ meson}

In this section we discuss the formalism for the study of the
$X(3872)$ meson. We adopt the convention that the spin and parity 
quantum numbers of the $X(3872)$ are $J^{PC} = 1^{++}$. Its mass
we express in terms of the binding energy $\epsilon_{\d\ds} > 0$ with
\eq 
m_X = m_{D^0} + m_{D^{\ast 0}} - \epsilon_{\d\ds} \,, 
\en 
where $m_{D^0} = 1864.85$ MeV and $m_{D^{\ast 0}} = 2006.7$ MeV 
are the $D^0$ and $D^{\ast 0}$ meson masses, respectively.

Following Ref.~\cite{Swanson:2003tb} we consider 
this state as superposition of the dominant molecular $\d\ds$ component and 
other hadronic configurations -- $D^{\pm} D^{\ast\,\mp}$, 
$\jp\omega$, and $\jp\rho$: 
\eq\label{Xstate}
|X(3872)\ra = \frac{Z_{\d\ds}^{1/2}}{\sqrt{2}}( |\d\dbs \ra + |\ds\db \ra )  
+ \frac{Z_{\dppm\dpmps}^{1/2}}{\sqrt{2}}( | \dpp\dpms \ra 
+ | \dpm\dpps \ra )  + Z_{\jpsi\omega}^{1/2} | \jpsi \omega \ra 
+ Z_{\jpsi\rho}^{1/2} | \jpsi \rho \ra 
\,, 
\en 
where $Z_{H_1H_2}$ is the probability to find the $X$ in the 
hadronic state $H_1H_2$ with the normalization 
$Z_{\d\ds} + Z_{\dppm\dpmps} + Z_{\jpsi\omega} +  Z_{\jpsi\rho} = 1$. 
For convenience, here and in the following we denote $J/\psi$ by $J_\psi$. 
The probabilities $Z_{H_1H_2}$ have been estimated in~\cite{Swanson:2003tb} 
as function of the binding energy $\epsilon$. 
Our approach is based on an effective interaction Lagrangian describing 
the couplings of the $X(3872)$ to its meson constituents. 
We apply two forms of such Lagrangians -- a local Lagrangian and
a nonlocal form containing the correlation functions 
$\Phi(y^2)$ characterizing the distribution of the constituents in the 
$X(3872)$). The simplest local Lagrangian reads   
\eq\label{Lagr_X_local}
{\cal L}_X^{\rm L}(x) &=& g_{_{X\d\ds}} \, X_\mu(x) \, J^\mu_{\d\ds}(x) 
+ g_{_{X\dppm\dpmps}} \, X_\mu(x) \, J^\mu_{\dppm\dpmps}(x) \nonumber\\
&+& \frac{g_{_{X\jpsi\omega}}}{m_X} \, 
\epsilon_{\mu\nu\alpha\beta} \, \partial^\nu X^\alpha(x) 
\, J^{\mu\beta}_{\jpsi\omega}(x) 
+ \frac{g_{_{X\jpsi\rho}}}{m_X} \, 
\epsilon_{\mu\nu\alpha\beta} \, \partial^\nu X^\alpha(x) 
\, J^{\mu\beta}_{\jpsi\rho}(x) 
\,,
\en 
where $g_{_{XH_1H_2}}$ is the coupling of $X(3872)$ to the constituents $H_1$ 
and $H_2$; $X$ is the field describing $X(3872)$; $J^\Gamma_{H_1H_2}$ 
is the current composed of the hadronic fields $H_1$ and $H_2$: 
\eq\label{Lagr_X_nonlocal} 
J^\mu_{D\bar D^\ast}(x) = \frac{1}{\sqrt{2}} (D(x)\bar D^{\ast\mu}(x) 
+ \bar D(x) D^{\ast\mu}(x))\,, 
\hspace*{.5cm} 
J^{\mu\beta}_{\jpsi V} = \jpsi^\mu V^\beta 
\en 
where $V=\rho, \omega$. 
 
The nonlocal version of the Lagrangian is obtained from the local one by 
introducing the correlation function into the hadronic 
current $J^\Gamma_{H_1H_2}$ as 
\eq\label{current_nonlocal} 
J^\mu_{D\bar D^\ast}(x) &\to& 
{\cal J}^\mu_{D\bar D^\ast}(x) = \frac{1}{\sqrt{2}} 
\int d^4y \Phi_{DD^\ast}(y^2) \biggl( D(x+y/2) \bar D^{\ast\mu}(x-y/2)  
+ \bar D(x+y/2) D^{\ast\mu}(x-y/2) \biggr) \,,\\ 
J^\mu_{\jpsi V}(x) &\to& 
{\cal J}^{\mu\beta}_{\jpsi V}(x) = \jpsi^{\beta}(x)  
\int d^4y \Phi_{V}(y^2) V^\mu(x+y) \,. 
\en 
Here $\Phi_{DD^\ast}$ is the correlation function describing the 
distribution of $D D^\ast$ inside $X$. The function $\Phi_{V}$ describes 
the distribution of the light vector meson $V=\rho$ or $\omega$ around 
the $\jpsi$, which is located at the center of mass of the $X(3872)$.
Since $m_V \ll m_{\jpsi}$ this description is like in heavy--light mesons
where the heavy quark $Q$
is surrounded by a light quark $q$ in the heavy quark limit of $m_q \ll m_Q$). 
A basic requirement for the choice of an explicit form of the correlation 
function $\Phi$ is that its Fourier transform vanishes sufficiently fast 
in the ultraviolet region of Euclidean space to render the Feynman diagrams 
ultraviolet finite. We adopt an identical Gaussian form
for both correlation functions $\Phi_{DD^\ast} = \Phi_{V} \equiv  \Phi_X$
in order to reduce the number of free parameters.
The Fourier transform
of the universal vertex function $\Phi_X$ is given by 
\eq 
\tilde\Phi_X(p_E^2/\Lambda^2) \doteq \exp( - p_E^2/\Lambda_X^2)\,,
\en 
where $p_{E}$ is the Euclidean Jacobi momentum. Here, $\Lambda_X$
is a size parameter. In Ref.~\cite{Dong:2008gb} the parameter was 
varied in the region 2 -- 3 GeV, a typical scales for 
$D$ and $D^\ast$ mesons - constituents of X(3872). 
In the present paper we fix the value to $\Lambda_X = 2$ GeV which is close  
to the masses of $D$ and $D^\ast$ mesons. One should remark, 
up to now we have no strong and direct justification for
the value of the $\Lambda_X$. The final conclusion about its magnitude
can done when we have more precise data on $X(3872)$.
Note, the local limit corresponds to the substitution of
$\Phi_X$ by the Dirac delta-function:
$\Phi_X(y^2) \to \delta^4 (y)$.

The coupling constants $g_{_{H_1H_2}}$ are determined by the compositeness 
condition~\cite{Weinberg:1962hj,Efimov:1993ei,Anikin:1995cf,%
Faessler:2007gv}. It implies that the renormalization constant of 
the hadron wave function is set equal to zero with 
\eq\label{ZX}
Z_X = 1 - \Sigma_X^\prime(m_X^2) = 0 \,.
\en
Here, $\Sigma^\prime_X(m_{X}^2) = d\Sigma_X(p^2)/dp^2|_{p^2=m_X^2}$ 
is the derivative of the transverse part of the mass operator 
$\Sigma^{\mu\nu}_X$, conventionally split into the transverse
$\Sigma_X$ and longitudinal $\Sigma^L_X$  parts as:
\eq
\Sigma^{\mu\nu}_X(p) = g^{\mu\nu}_\perp \Sigma_X(p^2) 
+ \frac{p^\mu p^\nu}{p^2} \Sigma^L_X(p^2) \,,
\en
where 
$g^{\mu\nu}_\perp = g^{\mu\nu} - p^\mu p^\nu/p^2$ 
and $g^{\mu\nu}_\perp p_\mu = 0\,.$ 
The mass operator of the $X(3872)$ receives contribution 
from four hadron--loop diagrams
\eq 
\Sigma_X(m_X^2) = \Sigma_{\d\ds}(m_X^2) + 
\Sigma_{\dppm\dpmps}(m_X^2) + \Sigma_{\jpsi\omega}(m_X^2) 
+ \Sigma_{\jpsi\rho}(m_X^2) \,. 
\en
induced by the  interaction of $X$ with the corresponding hadronic pairs 
$H_1H_2$ given in Eqs.~(\ref{Lagr_X_local}) and (\ref{Lagr_X_nonlocal}).
A typical diagram contributing to $\Sigma^{\mu\nu}_X(p)$ is shown in Fig.1.
Using Eq.~(\ref{Xstate}) and the compositeness condition~(\ref{ZX}) 
we get four independent equations to determine the 
coupling constants $g_{_{XH_1H_2}}$: 
\eq\label{ZH1H2} 
Z_{H_1H_2} = \Sigma_{H_1H_2}^\prime(m_X^2) \,. 
\en 
In order to evaluate the couplings $g_{_{XH_1H_2}}$ we use the
standard free propagators for the intermediate particles $H_1$ and $H_2$:
\eq 
iS_P(x-y)=\left<0|TP(x)P^\dagger (y)|0\right>=\int\frac{d^4k}{(2\pi)^4i}\, 
e^{-ik(x-y)} S_P(k),\quad S_P(k)=
\frac{1}{m_P^2-k^2-i\epsilon}
\en 
for pseudoscalar fields $P$ and 
\eq
iS_{V}^{\mu\nu}(x-y)=
\left<0|TV^\mu(x)V^{\nu\,\dagger} (y)|0\right>=
\int\frac{d^4k}{(2\pi)^4i}\,e^{-ik(x-y)} S^{\mu\nu}_V(k)\,,
\quad S^{\mu\nu}_V(k)=
\frac{-g^{\mu\nu}+k^\mu k^\nu/m_V^2}{m_V^2-k^2-i\epsilon}
\en 
for vector fields $V$.

Following Eqs.~(\ref{ZX}) and (\ref{ZH1H2}) in the nonlocal case the coupling
constants  
$g_{_{XH_1H_2}}$ are given by  
\eq\label{gXH1H2_couplings}
\frac{Z_{\d\ds}}{g_{_{X\d\ds}}^2} &=& \frac{1}{(4 \pi \Lambda_X)^2} \,
\int\limits_0^1 dx \int\limits_0^\infty
\frac{d\alpha \, \alpha \, P(\alpha, x)}{(1 + \alpha)^3} \, 
\Big( 1 + \frac{1}{4 \mu_{\ds}^2 (1 + \alpha)}\Big) \exp(z_1)\,, 
\label{gXDDS}\\ 
\frac{Z_{\jp V}}{g_{_{X\jp V}}^2} &=& \frac{1}{(4 \pi \Lambda_X)^2} \,
\int\limits_0^1 dx \int\limits_0^\infty
\frac{d\alpha \, \alpha \, Q(\alpha, x)}{(1 + \alpha)^3} \, 
\Big( 1 + \frac{1}{4 \mu_{\jpsi}^2 (1 + \alpha)}\Big) \exp(z_2)\,, \label{gXJV}
\en
where 
\eq 
& &P(\alpha, x) = \frac{\alpha}{2} \Big( 1 + 2 \alpha x (1-x) \Big)\,, \ \ \ 
   Q(\alpha, x) = \alpha x ( 1 + \alpha (1-x) )\,, \ \ \ 
   \mu_i = \frac{m_i}{\Lambda_X}\,, \nonumber\\ 
& &z_1 = - 2 \mu_{D^\ast}^2 \alpha x - 2 \mu_D^2 \alpha (1-x)
   + \frac{P(\alpha, x)}{1 + \alpha}  \, \mu_X^2 \,, \ \ \  
   z_2 = - 2 \mu_{\jp}^2 \alpha x - 2 \mu_V^2 \alpha (1-x) 
 + \frac{Q(\alpha, x)}{1 + \alpha}  \, \mu_X^2 \,. 
\en 
The expression for $g_{_{X\dppm\dpmps}}$ is obtained from (\ref{gXDDS}) 
by the corresponding replacement of masses and probability parameter 
$Z_{H_1H_2}$. 

In the local case we neglect the longitudinal part $k^\mu k^\nu/m_V^2$ 
of the vector meson propagator for the calculation of the coupling 
constants $g_{_{XH_1H2}}$ in order to have finite results. 
When writing the mass $m_H$ of the hadronic
molecule in the form $m_X = m_{H_1} + m_{H_2} - \epsilon_{H_1H_2}\,,$
where $\epsilon_{H_1H_1}$ represents the binding energy 
specific to a hadronic pair $(H_1H_2)$, we can perform
an expansion of $g_{_{XH_1H2}}^2$ in powers of $\epsilon_{H_1H_2}$. 
The leading-order
${\cal O}(\sqrt{\epsilon_{H_1H_2}})$ result of
\eq\label{eq:g}
\frac{g_{_{XH_1H2}}^2}{4\pi} = Z_{H_1H_2} \, C_{_{H_1H_2}} \, 
\frac{(m_{H_1}+m_{H_2})^{5/2}}{\sqrt{m_{H_1}m_{H_2}}} \, 
\sqrt{32\epsilon_{H_1H_2}}
\en
is in agreement with the ones derived in
Refs.~\cite{Weinberg:1962hj,Baru:2003qq,Branz:2008cb} 
also based on the compositeness condition $Z_X=0$. 
Here we have the factor $C_{_{H_1H_2}} = 1$ 
for $H_1H_2 = \d\ds, \dppm\dpmps$ and 
$C_{_{H_1H_2}} = 1/2$ for $\jpsi\omega, \jpsi\rho$. 

The numerical determination of the couplings $g_{_{XH_1H2}}$ for 
a specific hadron pair $H_1$ and $H_2$ shows that values obtained 
in the local and nonlocal case are very similar to each other. 
For example, for a binding energy of $\epsilon_{\d\ds} = 0.3$ MeV 
which corresponds to $m_X = 3.87151$ GeV, 
$\epsilon_{\dppm\dpmps} = 8.38$ MeV, 
$\epsilon_{\jpsi\omega} = 8.056$ MeV, 
and $\epsilon_{\jpsi\rho} = 0.896$ MeV we get in terms of 
the probability factors $Z_{H_1H_2}$ 
\eq 
& &g_{_{X\d\ds}} = 7.13 \ \text{GeV} \ \sqrt{Z_{\d\ds}} \ 
\text{(nonlocal)}\,, \hspace*{.25cm} 4.33 \ \text{GeV} 
\ \sqrt{Z_{\d\ds}} \ \text{(local)} \,, \nonumber\\ 
& &g_{_{X\dppm\dpmps}} = 11.39 \ \text{GeV} \ \sqrt{Z_{\dppm\dpmps}} \ 
\text{(nonlocal)}\,, \hspace*{.25cm} 9.98 \ \text{GeV} 
\ \sqrt{Z_{\dppm\dpmps}} \ \text{(local)} \,, \nonumber\\
& &g_{_{X\jpsi\omega}} = 6.59 \ \text{GeV} \ \sqrt{Z_{\jpsi\omega}} \ 
\text{(nonlocal)}\,, \hspace*{.25cm} 7.79 \ \text{GeV} 
\ \sqrt{Z_{\jpsi\omega}} \ \text{(local)} \,, \nonumber\\
& &g_{_{X\jpsi\rho}} = 4.93 \ \text{GeV} \ \sqrt{Z_{\jpsi\rho}} \ 
\text{(nonlocal)}\,, \hspace*{.25cm} 4.50 \ \text{GeV} \  
\sqrt{Z_{\jpsi\rho}} 
\ \ \text{(local)} \,.  
\en 
We point out that for the three couplings $g_{_{X\dppm\dpmps}}$, 
$g_{_{X\jpsi\omega}}$ and $g_{_{X\jpsi\rho}}$ there is no big 
difference between the nonlocal and local case. 
The reason is that 
the local couplings scale as $\epsilon_{H_1H_2}^{1/4}$. Therefore, 
a sizable deviation of the local coupling from the nonlocal one will 
only be relevant for values of $\epsilon_{H_1H_2} < 1$ MeV. For the nonlocal 
couplings the dependence on $\epsilon_{H_1H_2}$ is less pronounced. 
To illustrate this effect,
in Table 1 we indicate the dependence of $g_{_{X\d\ds}}/\sqrt{Z_{\d\ds}}$ 
as a function of $\epsilon_{\d\ds}$ both for the local and nonlocal case.
The nonlocal coupling 
changes slowly when $\epsilon_{\d\ds}$ is varied from 0.3 to 3 MeV. 
This is not the case for the local coupling: its value changes significantly 
when $\epsilon_{\d\ds}$ is increased from 0.3 to 1 MeV, but it remains more 
stable and gets closer to the result of the nonlocal case for 
$\epsilon_{\d\ds}\ge 1$ MeV.  
(This corresponds to the case of the other couplings
$g_{_{XH_1H_2}}$ calculated at $\epsilon_{H_1H_2} \ge 1$ MeV). 

\subsection{$X \to \chi_{cJ} + \pi^0$ transitions} 

In this subsection we consider the formalism for the two-body 
$X(3872) \to \chi_{cJ} + \pi^0$ transitions. Here 
the values of $J=0,1,2$ correspond to the $J^P=0^+, 1^+, 2^+$ 
quantum numbers of the charmonium states. The decays are described 
by the $(\d \ds)$ loop diagram shown in Fig.2. A further inclusion of 
the charged $(\dppm\dpmps)$ loops approximately gives the following 
correction to the decay widths 
\eq\label{Gamma_full}
\Gamma_0 \to \Gamma \simeq \Gamma_0 
\biggl(1 + \sqrt{\frac{Z_{X\dppm\dpmps}}{Z_{X\d\ds}}} \biggr)^2  \,. 
\en 
The diagrams of Fig.2 are generated by a phenomenological Lagrangian which 
contains two main parts: i) the first part is the 
Lagrangian derived in our approach describing the coupling of $X(3872)$ to
its constituents; ii) the second part is the set of interaction Lagrangians
describing the 
possible couplings of $D(D^\ast)$ mesons to pions and charmonia states. 
This second part can be taken from heavy hadron chiral perturbation theory  
(HHChPT)~\cite{Wise:1992hn,Jenkins:1992nb,Colangelo:2003sa}  
(for convenience we use a relativistic normalization of the meson states 
and write the Lagrangians in manifestly Lorentz covariant form): 
\eq 
{\cal L}_{D^\ast D \pi} &=& \frac{g_{D^\ast D \pi}}{\sqrt{2}} 
\biggl( D^\ast_{\mu i} i\partial^\mu \hat\pi_{ij} D^\dagger_j + 
{\rm H.c.}\biggr)  \,, \\
{\cal L}_{D^\ast D^\ast \pi} &=& \frac{g_{D^\ast D^\ast \pi}}{2\sqrt{2}} \, 
\epsilon_{\mu\nu\alpha\beta}
\biggl( D^{\ast\mu}_i \partial^\nu \hat\pi_{ij} 
\partial^\alpha D^{\ast\beta\dagger}_j + {\rm H.c.}\biggr)  \,, \\
{\cal L}_{\chi_{cJ}D^{(\ast)}D^{(\ast)}} &=& 
\chi_{c0} \biggl( g_{\chi_{c0}DD} D^\dagger_i D_i 
+ g_{\chi_{c0}D^\ast D^\ast} D^{\ast\dagger}_{\alpha i} 
D^{\ast \alpha}_i \biggr) \nonumber\\ 
&+& i g_{\chi_{c1}D^\ast D} \chi_{c1}^\alpha 
\biggl( D^\ast_{\alpha i} D^\dagger_i + \ {\rm H.c.} \ \biggr) 
+ g_{\chi_{c2}D^\ast D^\ast} \chi_{c2}^{\mu\nu} D^\ast_{\mu i} 
D^{\ast\dagger}_{\nu i} \,. 
\en 
Here $\hat\pi = \vec{\pi} \vec{\tau}$ is a $2 \times 2$ matrix containing 
the pion fields; $D$ and $D^\ast$ are the doublets of charm pseudoscalar 
and vector $D$ mesons; $\chi_{cJ}$ are the fields describing 
the charmonium states; $i,j$ are the isospin indices. The hadronic coupling
constants are expressed in terms of the universal HHChPT couplings $g, g_1$ 
and the hadronic masses as~\cite{Wise:1992hn,Jenkins:1992nb,Colangelo:2003sa} 
\eq 
& &g_{D^\ast D^\ast \pi} = \frac{g_{D^\ast D \pi}}
{\sqrt{m_{D} m_{D^\ast}}} = \frac{g}{F_\pi} \sqrt{2} 
\,, \nonumber\\
& &g_{\chi_{c0}DD} = 3 \frac{m_{D}}{m_{D^\ast}} 
g_{\chi_{c0}D^\ast D^\ast} 
= - 2 g_1 m_D \sqrt{3 m_{\chi_{c0}}} \,, \\ 
& &g_{\chi_{c1}D^\ast D} = g_1 \sqrt{2m_{\chi_{c1}} m_{D} m_{D^\ast}} \,, 
\nonumber\\ 
& &g_{\chi_{c2}D^\ast D} = 2 g_1 m_{D^\ast} \sqrt{m_{\chi_{c2}}} \,, 
\nonumber  
\en  
where $F_\pi = 92.4$ MeV is the leptonic decay constant. The coupling 
$g = 0.59$ (central value) is fixed from the data on the 
$\ds \to \d \pi$ branching 
ratio~\cite{Amsler:2008zz}. 
The coupling $g_1$ is related to the constant $f_{\chi_{c0}}$ 
parametrizing the matrix element 
$\la 0 | \bar c c | \chi_{c0}(p) \ra = f_{\chi_{c0}} 
m_{\chi_{c0}}$~\cite{Colangelo:2003sa} as 
\eq 
g_1 = - \sqrt{\frac{m_{\chi_{c0}}}{3}} \frac{1}{f_{\chi_{c0}}} \,. 
\en    
Using the estimate for $f_{\chi_{c0}} = 510$ MeV from
QCD sum rules~\cite{Colangelo:2002mj} 
we obtain for the coupling $g_1 = - 2.09$ GeV$^{-1/2}$. 

Evaluation of the diagrams in Fig.2 allows to write down an
effective Lagrangian corresponding to the
$X(3872) \to \chi_{cJ} \pi^0$ transitions with  
\eq
{\cal L}_{_{X\chi_{c0}\pi}} &=&
g_{_{X\chi_{c0}\pi}} \, X^\mu \, \partial_\mu\chi_{c0} \pi^0\,, 
\nonumber\\
{\cal L}_{_{X\chi_{c1}\pi}} &=& 
\frac{g_{_{X\chi_{c1}\pi}}}{m_X} \, 
\partial^\alpha X^\beta \, \chi_{c1}^\mu \, \partial^\nu\pi^0 \,  
\epsilon_{\mu\nu\alpha\beta}\,, \\
{\cal L}_{_{X\chi_{c2}\pi}} &=&
g_{_{X\chi_{c2}\pi}} \, X_\mu \, \chi_{c2}^{\mu\nu} 
\, \partial_\nu \pi^0 \,. 
\nonumber
\en
In terms of the effective couplings $g_{X\chi_{cJ}\pi}$
the decay widths of the $X(3872) \to \chi_{cJ} \pi^0$ transitions
are determined according to the expression:
\eq 
\Gamma(X(3872) \to \chi_{cJ} \pi^0) = 
\frac{P_\pi^2}{24\pi m_X^2} c_J g_{_{X\chi_{cJ}\pi}}^2 \, ,
\en 
where $c_J = 1$ for $J=0$  \ 2 for $J=1$ and
$5/3 (1 + 2 P_\pi^2/5m_{\chi_{c2}}^2)$ for $J=2$.   
Here $P_\pi = \lambda^{1/2}(m_X^2,m_{\chi_{cJ}}^2,m_\pi^2)/(2m_X)$ 
is the pion momentum in the $X(3872)$ rest frame and 
$\lambda(x,y,z) = x^2 + y^2 + z^2 - 2xy - 2yz - 2xz$ is the K\"allen 
function. 

\subsection{$X \to \chi_{cJ} + 2\pi$ transitions} 

For the three--body decays $X(3872) \to \chi_{cJ} + 2\pi^0$ we evaluate 
the diagrams of Fig.3. In our notation  $p$, $p_1$, 
$p_2$ and $p_3$ are the momenta of $X$, $\chi_{cJ}$ and the two pions, 
respectively. We introduce the invariant variables $s_i (i=1,2,3)$: 
\eq 
p   &=& p_1 + p_2 + p_3 \,, \nonumber\\
s_1 &=& (p_1 + p_2)^2 = (p - p_3)^2  \,, \nonumber\\
s_2 &=& (p_2 + p_3)^2 = (p - p_1)^2  \,, \\
s_3 &=& (p_1 + p_3)^2 = (p - p_2)^2  \,, \nonumber\\
s_1 + s_2 + s_3 &=& m_X^2 + m_{\chi_{cJ}}^2 + 2 m_\pi^2 \,. \nonumber
\en 
The decay widths are calculated according to the formula: 
\eq 
\Gamma(X(3872) \to \chi_{cJ} + 2\pi^0)&=&\frac{1}{1536 \pi^3 m_X^3} 
\int\limits_{4m_\pi^2}^{(m_X - m_{\chi_{cJ}})^2} ds_2 
\int\limits_{s_1^-}^{s_1^+} ds_1 \sum\limits_{\rm pol} |M_{\rm inv}|^2 \, ,
\en 
where 
\eq 
s_1^{\pm} &=& m_\pi^2 + \frac{1}{2} \biggl( m_X^2 + m_{\chi_{cJ}}^2 - s_2 
\pm \lambda^{1/2}(s_2,m_X^2,m_{\chi_{cJ}}^2) \sqrt{1 - \frac{4m_\pi^2}{s_2}}
\biggr)
\en
and $M_{\rm inv}$ is corresponding invariant matrix element. 

\subsection{Hadronic and radiative $X \rightarrow\jp + h$ decays} 

To get estimates for the decay widths of $X(3872) \to \jp + h$ with 
$h = \pi^+ \pi^- \pi^0, \pi^+ \pi^0, \pi^0 \gamma, \gamma$ 
we use the results of Ref.~\cite{Braaten:2005ai}, which are based on 
the assumption 
that these decays proceed through the processes $X$ to $\jp\omega$ 
and $\jp\rho$. 
In particular, it was shown that 
the $X(3872) \to \jp + h$ decay widths can be expressed in terms of 
$G_{_{X\jpsi V}}$ couplings as~\cite{Braaten:2005ai}: 
\eq\label{Xjpsidecays} 
& &\Gamma(X \to \jp \pi^+ \pi^-) 
= |G_{_{X\jpsi\rho}}|^2 
\cdot 223 \ 
\text{keV}\,, \nonumber\\ 
& &\Gamma(X \to \jp \pi^+ \pi^- \pi^0) = |G_{_{X\jpsi\omega}}|^2 \cdot 19.4 \ 
\text{keV}\,, \\ 
& &\Gamma(X \to \jp \pi^0 \gamma) \simeq 
|G_{_{X\jpsi\omega}}|^2 \cdot 3.24 \ \text{keV}\,, \nonumber\\ 
& &\Gamma(X \to \jp \gamma) = |G_{_{X\jpsi\rho}} 
+ 0.30 G_{_{X\jpsi\omega}}|^2 
\cdot 5.51 \ \text{keV}\,.  \nonumber  
\en  
The couplings $G_{X\jpsi V}$ introduced in~\cite{Braaten:2005ai} are 
related to our set of couplings $g_{X\jpsi V}$ as: 
\eq\label{X-rel}
G_{_{X\jpsi V}} = \frac{g_{_{X\jpsi V}}}{m_V} \,. 
\en
In our approach, based on the representation (\ref{Xstate}) for the $X$, 
we deduced the effective couplings $g_{_{X\jpsi\omega}}$ and 
$g_{_{X\jpsi\rho}}$ in terms of the unknown probabilities 
$Z_{X\jpsi\omega}$ and $Z_{X\jpsi\rho}$. These 
results we use in Eqs.   (\ref{Xjpsidecays})-(\ref{X-rel}).
Note that Eqs.  (\ref{Xjpsidecays})-(\ref{X-rel}),
corresponding to the $X \to \jp + h$ decays,
only take into account short--distance effects~\cite{Braaten:2005ai}. 
To be consistent one should also include long--distance effects 
due to the contribution of the molecular $\d\ds$ component. Such a 
detailed analysis goes beyond the scope of the present work. 
Here we estimate both short and long-distances effects only for the 
$X \to \gamma \jp$ decays using our previous results on the molecular 
contribution obtained in Ref.~\cite{Dong:2008gb}.  

\section{Results} 

We present our numerical results in terms of the probabilities 
$Z_{H_1H_2}$ and then substitute the typical values for 
$Z_{H_1H_2}$ based on the estimate of Ref.~\cite{Swanson:2003tb} for
a binding energy of
$\epsilon = 0.3$ MeV: 
\eq\label{ZH1H2_factors} 
Z_{\d\ds} = 0.92\,, \hspace*{.25cm} 
Z_{\dppm\dpmps} = 0.033\,, \hspace*{.25cm} 
Z_{\jpsi\omega} = 0.041\,, \hspace*{.25cm} 
Z_{\jpsi\rho} = 0.006\,. 
\en  
In Table 2 we present  our results for the $X \to \chi_{cJ} + \pi^0 (2 \pi^0)$ 
decay widths and the ratios 
\eq 
R_{cJ} = \frac{\Gamma(X \to \chi_{cJ} + 2 \pi^0)} 
              {\Gamma(X \to \chi_{cJ} +\pi^0)} \,. 
\en 
We also give predictions for 
the effective couplings $g_{_{X\chi_{cJ}\pi}}$. In the second column 
we indicate the contribution of the $\d\ds$ loop only.
Results are given in terms of the $Z_{H_1H_2}$ factors and
values in brackets are based on the explicit numbers of
Eq.~(\ref{ZH1H2_factors}).
The third column contains the results 
including both $\d\ds + \dpm\dpps$ loops, again based on
the probability factors of Eq.~(\ref{ZH1H2_factors}).
In the fourth column we give the predictions based on the approximate 
formula~(\ref{Gamma_full}). We also introduce the notation 
$\beta = (Z_{\dppm\dpmps}/Z_{\d\ds})^{1/2}$ for the ratio of
the probability factors. Again, values in brackets are deduced
with the explicit values for $Z_{H_1H_2}$. 

The $\d\ds$ molecular component gives (as naively expected)
the dominant contribution to the $X \to \chi_{cJ} + \pi^0, 2 \pi^0$ 
rates. Also, the results based on the approximate 
expression~(\ref{Gamma_full}) including the charged
$\dppm\dpmps$ component turn out to be quite close to
the exact calculation. 
Comparing our predicted ratios of Table 2 to the results of 
Ref.~\cite{Fleming:2008yn} 
\eq 
R_{c0} = 9.1 \times 10^{-6}\,, \hspace*{.5cm}
R_{c1} = 6.1 \times 10^{-1}\,, \hspace*{.5cm}
R_{c2} = 7.8 \times 10^{-6}\,
\en 
larger differences occur. 
This is especially due to the nonrelativistic treatment of the $\d$
and $\ds$ mesons in Ref.~\cite{Fleming:2008yn}. 
The large value of $R_{c1}$ in Ref.~\cite{Fleming:2008yn} is 
sensitive to the treatment of the pole position of the nonrelativistic 
energy denominator and to the width of the $D^0$ meson. 

In Table 3 we present our results for the $X \to \jp + h$ 
decays as based on the set of relations of Eq.~(\ref{Xjpsidecays}).
The predictions are given both for the local and nonlocal
cases.
Again, final results are given in terms of the relevant
$Z_{H_1H_2}$ factors, using in addition the notation
$\sigma = (Z_{\jpsi\rho}/Z_{\jpsi\omega})^{1/2}$, while
numbers in brackets are based on Eq.~(\ref{ZH1H2_factors}).
For the probability factors of Eq.~(\ref{ZH1H2_factors}) we
also list our results for the ratios 
\eq  
R_1 = \frac{\Gamma(X \to \jp \pi^+ \pi^- \pi^0)}
{\Gamma(X \to \jp \pi^+ \pi^-)}\,, \hspace*{.5cm}
R_2 = \frac{\Gamma(X \to \jp \gamma)}
{\Gamma(X \to \jp \pi^+ \pi^-)} \,,
\en 
related to the present experimental
situation given in Eqs.~(\ref{data_strong}) and (\ref{data_radiative}).
One can see, that nonlocal and local cases are numerically
similar to each other. 
To our mind only the decay width $\Gamma(X \to \jp \pi^+ \pi^- \pi^0)$
and hence the ratio $R_1$ might be 
overestimated in the local case.
Also note that the results for $R_1$ and $R_2$ in
the more realistic, nonlocal case are consistent with
present experimental findings displayed in Eqs.~(\ref{data_strong})
and (\ref{data_radiative}). Let us remark that the results obtained in 
the local case are close to the nonlocal case. As one can from the numbers, 
the local approximation including
truncation of the vector meson propagator is reasonable approximation to
the nonlocal case at $\Lambda_X = 2$ GeV. When $\Lambda_X$ is
increasing the difference of two cases becomes more sizable.
 
Next we also want to comment on the result for the
decay width $\Gamma(X \to \jp \gamma)$. 
In Ref.~\cite{Dong:2008gb} we originally gave an estimate for this decay width
including the molecular 
$\d\ds$ and the $c\bar c$ charmonium components.
We showed that the contribution of 
the charmonium component is strongly suppressed. For a cutoff value of
$\Lambda = 2$ GeV our result 
for $\Gamma(X \to \jp \gamma)$ was 118.9 keV. 
In Ref.~\cite{Dong:2008gb} we 
described the couplings of  
$\jp$ to $\d\d$ and $\ds\ds$ applying a phenomenological Lagrangian 
used in the analysis of $J/\psi$~\cite{Lin:1999ad}. 
We also did not include possible, additional form factors
at the meson interaction vertices for reasons of simplicity and in order 
to have less free parameters. Inclusion of such form factors could lead 
to a further reduction of the predicted value for the $X \to \jp \gamma$ 
decay width. The importance of these form factors was recognized before 
in connection with different aspects of charm physics, in particular, 
with the suppression of  the $J/\psi$
dissociation cross sections~\cite{Matinyan:1998cb}. 
This implies that our result of Ref.~\cite{Dong:2008gb} corresponds to an
upper limit for the decay width
$\Gamma(X \to \gamma \jp)$. Let us note that this value can be further 
reduced by the following four effects: i) by the probability factor $Z_{\d\ds}$; 
ii) when using smaller values for the couplings of $\jp$ to
the $\d\d$ and $\ds\ds$ 
pairs (in Ref.~\cite{Dong:2008gb} we used 
$g_{_{J_\psi DD}} = g_{_{J_\psi D^\ast D^\ast}} = 6.5$);  
iii) by the inclusion of form factors in the $\jp\d\d$ and 
$\jp\ds\ds$ vertices;  
iv) when taking into account the short--distance mechanism of the 
$X \to \jp + V [\to \gamma]$ transition, considered presently,
leading to destructive interference with the molecular 
contribution. Without introducing form factors at the
$\jp\d\d$ and $\jp\ds\ds$ vertices 
and taking into account three additional suppression effects [i), ii) and iv)]
we now have for 
$\Gamma(X \to \jp \gamma)$ in terms of the coupling 
$g_{J_\psi} = g_{_{J_\psi DD}} =g_{_{J_\psi D^\ast D^\ast}}$: 
\eq 
\Gamma(X \to \jp \gamma) = (1.605 \, g_{J_\psi} - 2.354)^2 \ \text{keV} \,. 
\en   
When varying $g_{J_\psi}$ from 5 to 6.5 we get 
\eq 
\Gamma(X \to \jp \gamma) = 32.2 - 65.3 \ \text{keV} \, , 
\en 
where a further possible reduction of this value can be obtained
by including form factors at the
$\jp\d\d$ and $\jp\ds\ds$ vertices. Note, that three different results for the 
$\Gamma(X \to \jp \gamma)$ are obtained using different approximation 
for the $X(3872)$ wave function: 
i) 64.4 - 118.9 keV was obtained for a mixture of molecular $DD^\ast$
and charmonium $c\bar{c}$ components;
ii) 5.5 keV was obtained for pure $J/\psi V$ components;
iii) 32.2 - 65.3 keV was obtained taking a destructive interference 
of molecular $DD^\ast$ and charmonium $c\bar{c}$ components with 
$J/\psi V$ components. 

Our final comment concerns the
$X \to \psi(2s) + \gamma$ decay width recently 
measured by the {\it BABAR} Collaboration~\cite{Aubert:2008rn}:
\eq\label{data_radiative_psi2}
R_3 = \displaystyle\frac{\Gamma(X\to \psi(2s) \gamma)}
{\Gamma(X\to \jp \gamma)} = 3.5 \pm 1.4 
\en 
In our opinion this value can be interpreted as a signal 
for mixing of the $\d\ds$ and $\jp V$ components  
in the $X\to \jp \gamma$ mode. In the $X\to \psi(2s) \gamma$ 
transition only the molecular $\d\ds$ component will contribute
under the condition that a $\psi(2s) V$ component 
in the $X(3872)$ is completely absent or suppressed relative to
the $\jp V$ configurations. In the future we plan to calculate 
all the decay modes $X \to \jp h$ including $X \to \psi(2s) \gamma$ 
using the HHChPT Lagrangian~\cite{Colangelo:2003sa}.  

\section{Summary}

We have considered the $X(3872)$ resonance with 
$J^{PC} = 1^{++}$ as a composite hadronic state made up of a  
dominant molecular $\d\ds$ component and other
hadronic pairs -- $D^{\pm} D^{\ast\,\mp}$, 
$\jp\omega$ and $\jp\rho$. Applying the compositeness condition 
we constrained the couplings of $X(3872)$ to its constituents. 
We calculated two- and three-body hadronic decays of the $X(3872)$ 
to charmonium states $\chi_{cJ}$ and pions using a phenomenological 
Lagrangian approach. Then using the estimated $X\jp\omega$ and $X\jp\rho$ 
couplings we calculated the widths of $X(3872) \to \jp + h$ transitions, 
where $h$ = $\pi^+\pi^-$, $\pi^+\pi^-\pi^0$, $\pi^0\gamma$ and $\gamma$.
The full, structure-dependent decay pattern of the $X(3872)$ developed here
can serve to possibly identify its hadronic composition in
running and planned experiments. 

\newpage 

\begin{acknowledgments}

This work was supported by the DFG under Contract No. FA67/31-1,
No. FA67/31-2, and No. GRK683.
This work is supported  by the National Sciences Foundations 
No. 10775148 and by CAS Grant No. KJCX3-SYW-N2 (YBD). 
This research is also part of the
European Community-Research Infrastructure Integrating Activity
``Study of Strongly Interacting Matter'' (HadronPhysics2,
Grant Agreement No. 227431) and of the President grant of Russia
``Scientific Schools''  No. 871.2008.2.
Y.B.D. would like to thank the T\"ubingen theory group for 
for its hospitality. 
V.E.L. would like to thank the theory group of Universidad T\'ecnica 
Federico Santa Mar\'\i a for its hospitality. 
This work was partially supported by the PBCT Project No. ACT-028 
Center of Subatomic Physics.
\end{acknowledgments}

\begin{figure}[htb]

\vspace*{2cm} 

\begin{center}
\epsfig{file=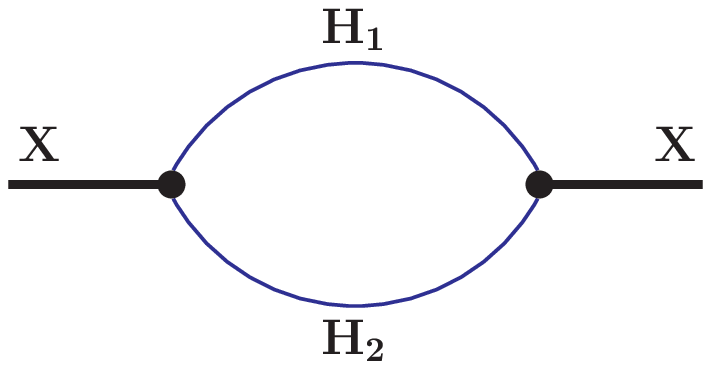, scale=.75}
\end{center}
\caption{$H_1H_2$ hadron--loop diagrams contributing to the mass operator 
of the $X(3872)$ meson.}

\vspace*{2cm}

\begin{center}
\epsfig{file=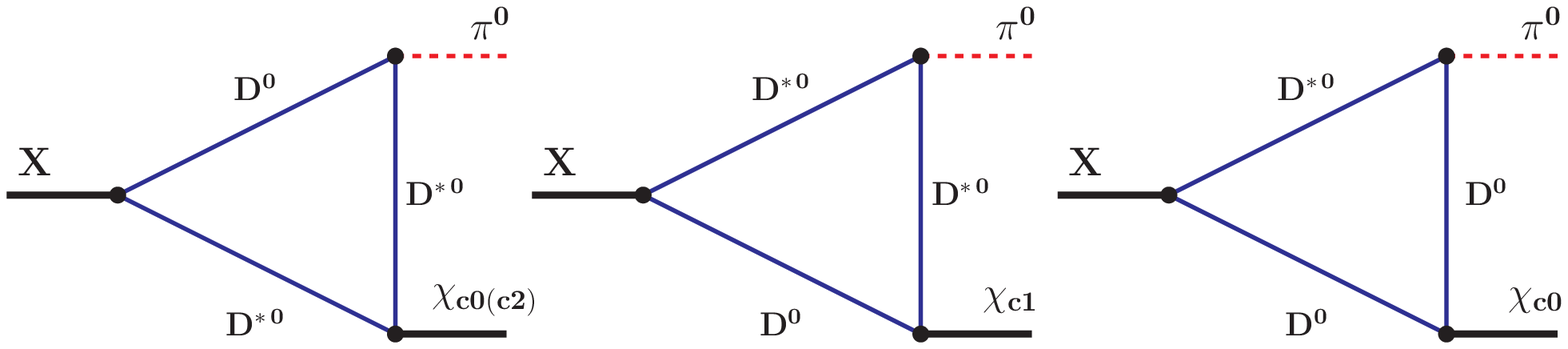, scale=.75}
\end{center}
\caption{Diagrams contributing to the hadronic transitions 
$X(3872) \to \chi_{cJ} + \pi^0$.}
\end{figure}

\newpage

\begin{figure}
\begin{center} 
\epsfig{file=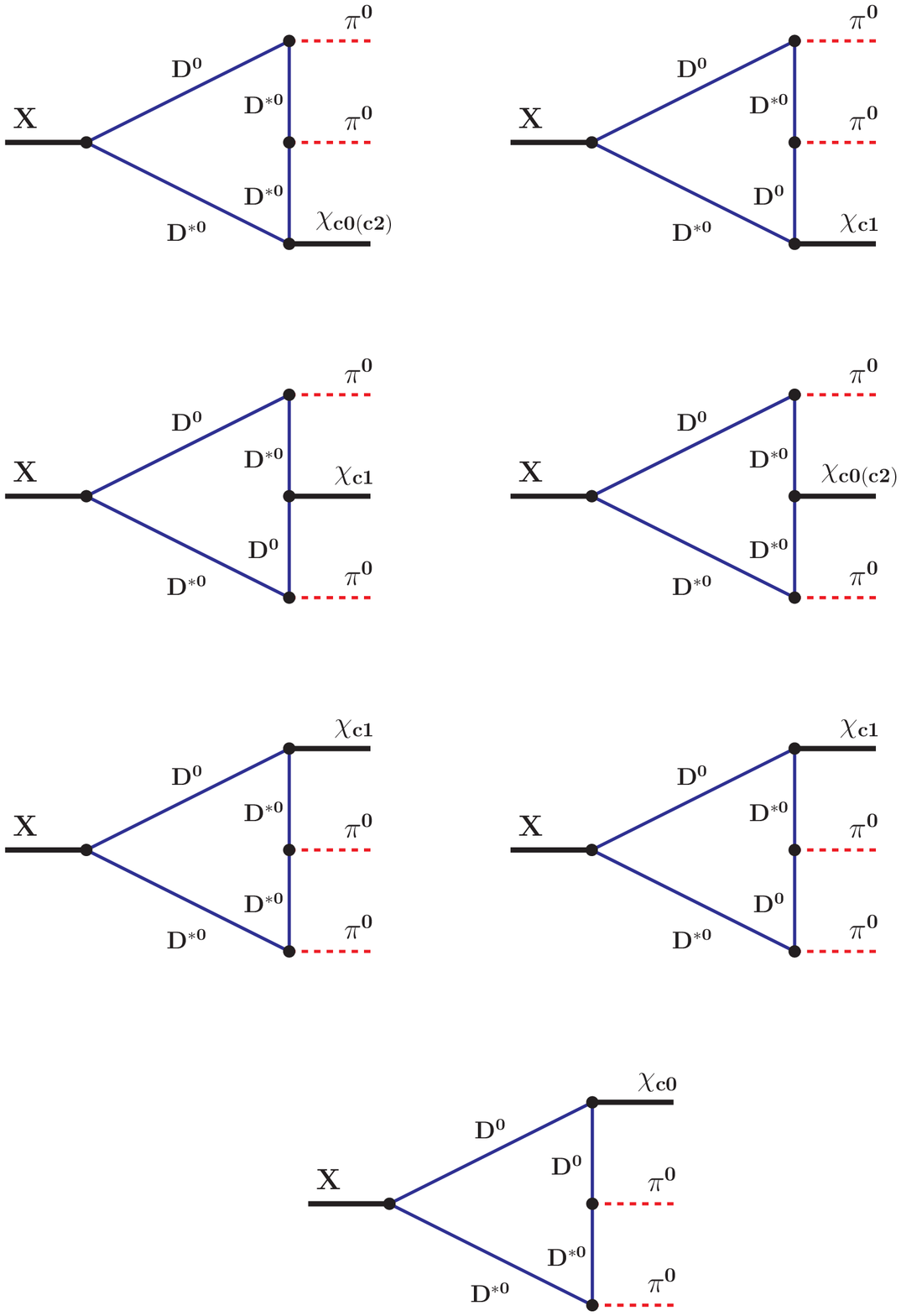, scale=.75}
\end{center}
\caption{Diagrams contributing to the hadronic transitions
$X(3872) \to \chi_{cJ} + 2 \pi^0$.}
\end{figure}

\newpage 

\begin{table}[htb]
\begin{center}
{\bf Table 1.} 
Dependence of the couplings $G_{_{X\d\ds}} = g_{_{X\d\ds}}/Z_{\d\ds}^{1/2}$ 
on the binding energy $\epsilon_{\d\ds}$. 

\vspace*{.25cm}

\def\arraystretch{2}
\begin{tabular}{|l|l|l|l|l|l|l|l|l|l|l|l|l|}  
\hline
\hspace*{1cm} $\epsilon_{\d\ds}$ (MeV) 
& 0.3 & 0.5 & 0.7 & 1 & 1.3 & 1.5 & 1.7 & 2 & 2.3 & 2.5 & 2.7 & 3 \\
\hline
Local case: \hspace*{.3cm} 
$G_{_{X\d\ds}}$ (GeV) & 
4.33 & 4.92 & 5.35 & 5.85 & 6.25 & 6.48 & 6.69 & 6.96 & 7.21 & 
7.36 & 7.50 & 7.70 \\ 
\hline 
Nonlocal case: $G_{_{X\d\ds}}$ (GeV) & 
7.13 & 7.25 & 7.37 & 7.54 & 7.72 & 7.83 & 7.94 & 8.11 & 8.28 & 
8.39 & 8.49 & 8.65 \\
\hline
\end{tabular}
\end{center}

\vspace*{.4cm}

\begin{center}
{\bf Table 2.} 
Properties of $X \to \chi_{cJ} + \pi^0 (2 \pi^0)$ decays.
The numbers in brackets and for column \\ $\d\ds + \dpm\dpps$ [exact]
result from explicit values for $Z_{\d\ds}$ and  
$\beta = (Z_{\dppm\dpmps}/Z_{\d\ds})^{1/2}$ of Eq.~(\ref{ZH1H2_factors}).

\vspace*{.25cm}
\def\arraystretch{2}
\begin{tabular}{|l|l|l|l|} \hline
\hspace*{.7cm} Quantity & \hspace*{.8cm} $\d\ds$ loop 
& \hspace*{.05cm} $\d\ds + \dpm\dpps$   
& \hspace*{1cm} $\d\ds + \dpm\dpps$ \\
      &
      & \hspace*{.75cm} [exact] & \hspace*{1.5cm} [Eq.~(\ref{Gamma_full})] \\ 
\hline 
$g_{_{X\chi_{c0}\pi}}$ &0.826 $\sqrt{Z_{\d\ds}}$(0.792) 
                       &1.007 
                       &0.826 $\sqrt{Z_{\d\ds}}(1+\beta)$(0.942) 
\\ \hline 
$g_{_{X\chi_{c1}\pi}}$ &0.444 $\sqrt{Z_{\d\ds}}$(0.426) 
                       &0.539
                       &0.444 $\sqrt{Z_{\d\ds}}(1+\beta)$(0.507)  
\\ \hline 
$g_{_{X\chi_{c2}\pi}}$ &0.655 $\sqrt{Z_{\d\ds}}$(0.628) 
                       &0.797
                       &0.655 $\sqrt{Z_{\d\ds}}(1+\beta)$(0.747)  
\\ \hline 
$\Gamma(X \to \chi_{c0} + \pi^0)$, keV & 41.1 $Z_{\d\ds}$ (37.8)
                                       & 61.0
                                       & 41.1  $Z_{\d\ds}$ \, 
                                       $(1 + \beta)^2$ (53.5) 
\\ \hline 
$\Gamma(X \to \chi_{c0} + 2\pi^0)$, eV & 63.3 $Z_{\d\ds}$ (58.2)
                                       & 94.0 
                                       & 63.3 $Z_{\d\ds}$ \, 
                                       $(1 + \beta)^2$ (82.4) \\ 
\hline 
$R_{c0} \times 10^3$ &1.54  &1.54 & 1.54 \\ \hline 
$\Gamma(X \to \chi_{c1} + \pi^0)$, keV & 11.1 $Z_{\d\ds}$ (10.2)
                                       & 16.4 
                                       & 11.1 $Z_{\d\ds}$ \,  
                                       $(1 + \beta)^2$ (14.5) \\ 
\hline 
$\Gamma(X \to \chi_{c1} + 2\pi^0)$, eV & 743 $Z_{\d\ds}$ (683.6)
                                       & 1095.2
                                       & 743 $Z_{\d\ds}$ \, 
                                       $(1 + \beta)^2$ (969.6) \\ 
\hline 
$R_{c1} \times 10^2$ & 6.69 &6.68 & 6.69 \\ \hline 
$\Gamma(X \to \chi_{c2} + \pi^0)$, keV & 15 $Z_{\d\ds}$ (13.8)
                                     & 22.1 
                                     & 15 $Z_{\d\ds}$ \, 
                                     $(1 + \beta)^2$ (19.5) \\ 
\hline 
$\Gamma(X \to \chi_{c2} + 2\pi^0)$, eV & 20.6 $Z_{\d\ds}$ (19.0)
                                       & 30.4
                                       & 20.6 $Z_{\d\ds}$ \, 
                                       $(1 + \beta)^2$ (26.9) \\ 
\hline 
$R_{c2} \times 10^3$ & 1.38 &1.38 & 1.38 \\
\hline 
\end{tabular}
\end{center}

\vspace*{.5cm}

\begin{center}
{\bf Table 3.} 
Properties of $X \to \jpsi + h$ decays.  
The numbers in brackets and for the ratios $R_1$, $R_2$ \\ 
from explicit values for $Z_{\jpsi\rho}$, $Z_{\jpsi\omega}$ and  
$\sigma = (Z_{\jpsi\rho}/Z_{\jpsi\omega})^{1/2}$
of Eq.~(\ref{ZH1H2_factors}).

\vspace*{.25cm}

\def\arraystretch{2}
\begin{tabular}{|l|l|l|}  
\hline
\hspace*{1cm} 
Quantity & 
\hspace*{1cm} 
Local case & 
\hspace*{1cm} 
Nonlocal case \\ 
\hline 
$\Gamma(X \to \jp \pi^+ \pi^-)$, keV       
& $7.5 \times 10^3  \, Z_{\jpsi\rho}$ (45.0)
                                           
& $9.0 \times 10^3  \, Z_{\jpsi\rho}$ (54.0) \\
\hline
$\Gamma(X \to \jp \pi^+ \pi^- \pi^0)$, keV 
& $1.92 \times 10^3  \, Z_{\jpsi\omega}$ (78.9)
                                           
& $1.38 \times 10^3  \, Z_{\jpsi\omega}$ (56.6) \\
\hline
$\Gamma(X \to \jp \pi^0 \gamma)$, keV      
& $0.32 \times 10^3  \, Z_{\jpsi\omega}$ (13.2)
                                           
& $0.23 \times 10^3  \, Z_{\jpsi\omega}$ (9.4) \\ 
\hline 
$\Gamma(X \to \jp \gamma)$, keV            
& $49.18 \, Z_{\jpsi\omega} \, (1 + 1.94 \sigma)^2$ (6.1)
                                           
& $35.19 \, Z_{\jpsi\omega} \, (1 + 2.51 \sigma)^2$ (5.5)\\
\hline 
$R_1$ & 1.75 & 1.05 \\
\hline
$R_2$ & 0.14 & 0.10 \\
\hline
\end{tabular}
\end{center}
\end{table}

\end{document}